%% file: main.tex
\documentclass[manuscript,screen]{acmart}
\usepackage{placeins}

\AtBeginDocument{%
  }

\setcopyright{acmlicensed}
\copyrightyear{2018}
\acmYear{2018}
\acmDOI{XXXXXXX.XXXXXXX}

\acmConference[Conference acronym 'XX]{Make sure to enter the correct
  conference title from your rights confirmation emai}{June 03--05,
  2018}{Woodstock, NY}
\acmISBN{978-1-4503-XXXX-X/18/06}

\input{sections/header}
\begin{document}


\title{
\sys{}: Understanding AI Usage by Visualizing Student-AI Interaction in Code
}


\input{sections/authors}

\renewcommand{\shortauthors}{Zhang et al.}

\input{sections/00-abstract}

\ccsdesc[500]{Human-centered computing~Interactive systems and tools}
\ccsdesc[500]{Human-centered computing~Information visualization}
\ccsdesc[500]{Applied computing~Education}
\keywords{Programming Education; Code Visualization; Student-AI Interaction}


\received{20 February 2007}
\received[revised]{12 March 2009}
\received[accepted]{5 June 2009}

\maketitle

\input{sections/document}

\bibliographystyle{ACM-Reference-Format}
\bibliography{refs}

\input{sections/09-appendix}

\end{document}

%% file: sections/header.tex
\input{commands-packages/dev-packages}
\input{commands-packages/dev-commands}

\input{commands-packages/publish-packages}
\input{commands-packages/publish-commands}
\input{commands-packages/publish-aliases}

\input{figures/figures}
\input{tables/tables}

%% file: commands-packages/dev-packages.tex
\usepackage{xargs}      
\usepackage{soul}       
\usepackage{color}      
\usepackage{graphicx}  
\usepackage{amsmath}
\usepackage{minted}

\newcommand{\added}[2][]{#2}      
\newcommand{\deleted}[2][]{}
\newcommand{\replaced}[3][]{#2}   


%% file: commands-packages/dev-commands.tex
\definecolor{LIGHTPINK}{RGB}{237,157,202}
\definecolor{LIGHTRED}{RGB}{210,121,121}
\definecolor{LIGHTORANGE}{RGB}{230,170,50}
\definecolor{LIGHTGOLD}{RGB}{210,194,121}
\definecolor{LIGHTGREEN}{RGB}{121,210,121}
\definecolor{LIGHTAQUA}{RGB}{121,206,210}
\definecolor{LIGHTBLUE}{RGB}{121,124,210}
\definecolor{LIGHTPURPLE}{RGB}{153,102,255}
\definecolor{RED}{RGB}{178,34,34}
\definecolor{GRAY}{RGB}{166,166,166}
\definecolor{WHITE}{RGB}{255,255,255}
\definecolor{DARKGREEN}{rgb}{0.0, 0.5, 0.0}
\definecolor{SENSAICHA}{HTML}{4D5139}

\newcommandx{\guest}[3][1=]
    {\setulcolor{LIGHTORANGE}{\ul{#1}} \textcolor{LIGHTORANGE} 
    {[\textbf{#2:} #3]}}

\newcommandx{\gez}[2][1=]
    {\setulcolor{LIGHTBLUE}{\ul{#1}} \textcolor{LIGHTBLUE}{[\textbf{Ashley:} #2]}}

\newcommandx{\soney}[2][1=]
    {\setulcolor{LIGHTPURPLE}{\ul{#1}} \textcolor{LIGHTPURPLE}{[\textbf{Steve:} #2]}}

\newcommandx{\yanru}[2][1=]
    {\setulcolor{DARKGREEN}{\ul{#1}} \textcolor{DARKGREEN}{[\textbf{Yanru:} #2]}}



%% file: commands-packages/publish-packages.tex
\usepackage{xcolor}
\usepackage{booktabs} 
\usepackage{tikz}

\usepackage{booktabs}
\usepackage{multirow}
\usepackage{graphicx}
\usepackage{array}      
\usepackage{makecell}   

%% file: commands-packages/publish-commands.tex
\newcommand{\myquote}[1]{\emph{``#1''}} 

\newcommand{\mylongquote}[1]{           
    \begin{quote}\emph{#1}\end{quote}}

\newcommand{\dgbadge}[2][teal]{%
  \tikz[baseline=(n.base)]\node
    [fill=#1!85!black, rounded corners=2pt,
     inner sep=2pt, outer sep=0pt,
     text=white, font=\bfseries\footnotesize] (n){#2};%
}



%% file: commands-packages/publish-aliases.tex
\newcommand{\sys}{Editrail}


%% file: figures/figures.tex
\newcommand{\figureSys}{
\begin{figure*}
    \centering
    \includegraphics[width=\linewidth]{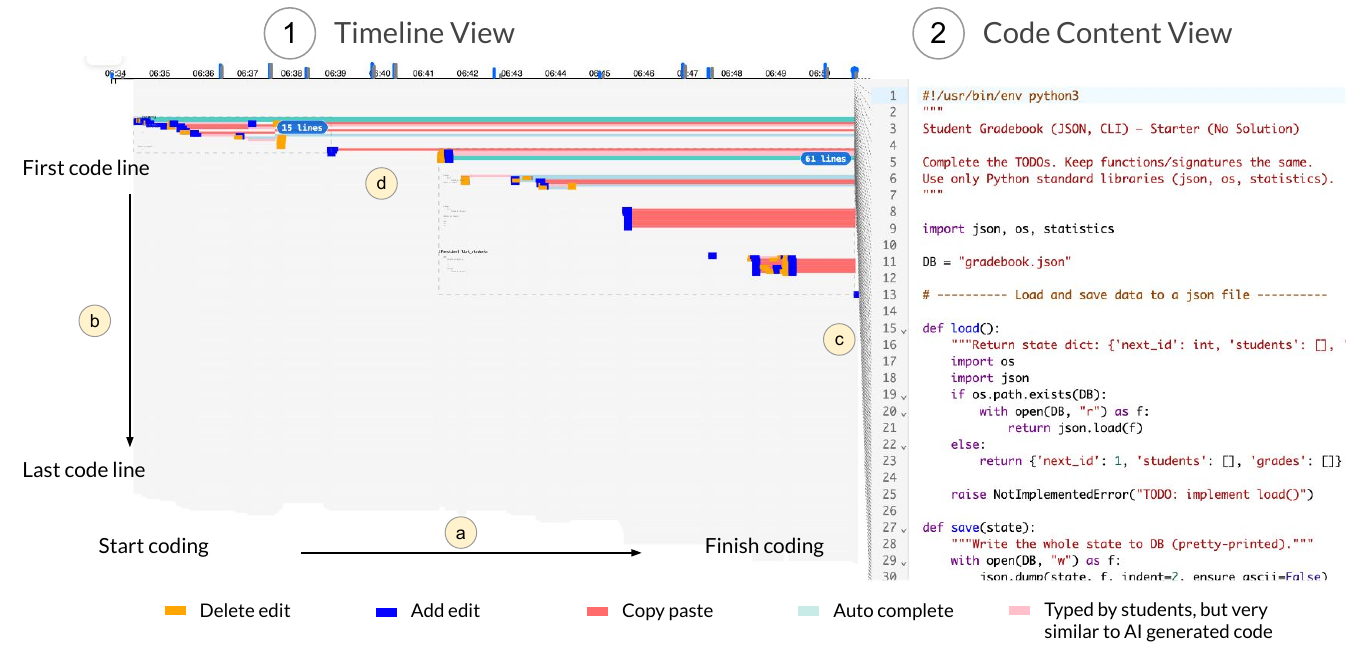}
    \caption{User interface of \sys{}. (1) Timeline View: rows represent code lines and time flows horizontally. Edits are shown as colored overlays indicating different types of AI involvement (d): red for copy–paste, green for autocomplete, and pink for student-typed code that closely resembles AI-generated output. Indicators also show coding progress from start to finish. (2) Code Content View: aligned program text showing the full source code at the selected point in time, enabling instructors to connect timeline annotations with specific code states.}
    \label{fig:sys}
\end{figure*}
}

\newcommand{\figureUIDetailsTimeline}{
\begin{figure*}
    \centering
    \includegraphics[width=0.5\textwidth]{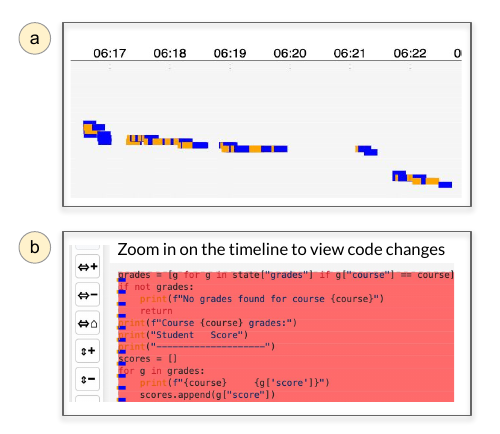}
    \caption{Details of \sys{}’s timeline interaction. (a) Keystroke-level edit indicators shown along the timeline, where blue ones represent insertion and orange ones represent deletion edits. (b) Zoomed-in timeline view revealing the exact code content associated with individual change indicators for detailed inspection. }
    \label{fig:ui-details-timeline}
\end{figure*}
}

\newcommand{\figureUIDetailsAI}{
\begin{figure*}
    \centering
    \includegraphics[width=\textwidth]{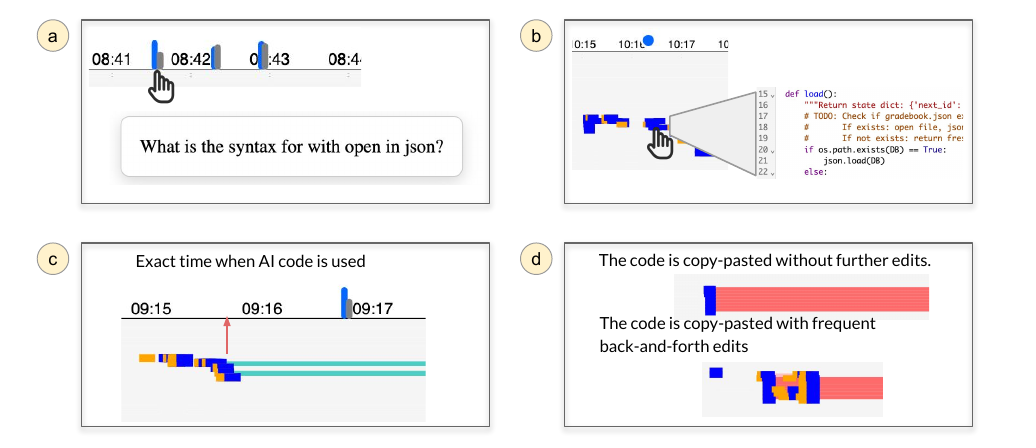}
    \caption{Details of \sys{}'s interaction of monitoring AI usage. (a) AI–student dialogue shown at the exact point of interaction. (b) Timeline-to-code projection aligning edits with program text. (c) Precise timing of AI code usage. (d) Code pasted from AI without further edits and code pasted from AI with frequent back-and-forth edits by the student.}
    \label{fig:ui-details-ai}
\end{figure*}
}

\newcommand{\figureQA}{
\begin{figure*}
    \centering
    \includegraphics[width=\linewidth]{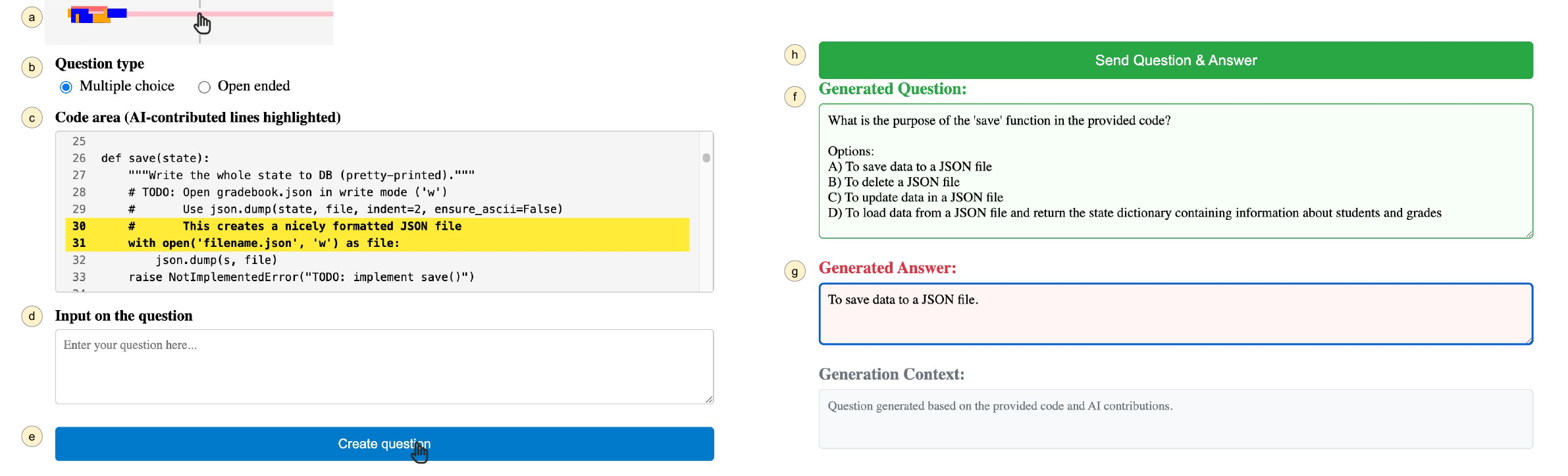}
    \caption{User interface of \sys{}'s question creation. (a) Instructors select a point on the coding timeline to create questions. (b) Question type options (multiple choice or open-ended). (c) Code area highlighting AI-contributed lines in student's code. (d) Input field for customizing the question. (e) Button to generate the question. (f) Example of a system-generated question based on the selected code. (g) Generated answer for the question. (h) Send the question to students.}
    \label{fig:ui-qa}
\end{figure*}
}

\newcommand{\figureAISources}{
\begin{figure*}
    \centering
    \includegraphics[width=\linewidth]{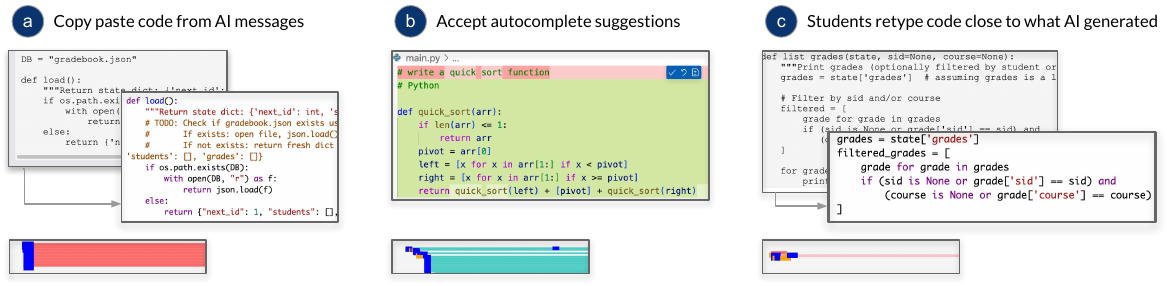}
    \caption{Three cases of code that originates from AI. (a) the user copies and pastes code directly from an AI agent (e.g., ChatGPT or Copilot chat). (b) the user accepts an autocomplete suggestion (e.g., from Copilot's in-editor hints). (c) The user types code that is very similar to something that was part of their conversation with an AI agent.}
    \label{fig:ai-sources}
\end{figure*}
}

\newcommand{\figureBaseline}{
\begin{figure*}
    \centering
    \includegraphics[width=\linewidth]{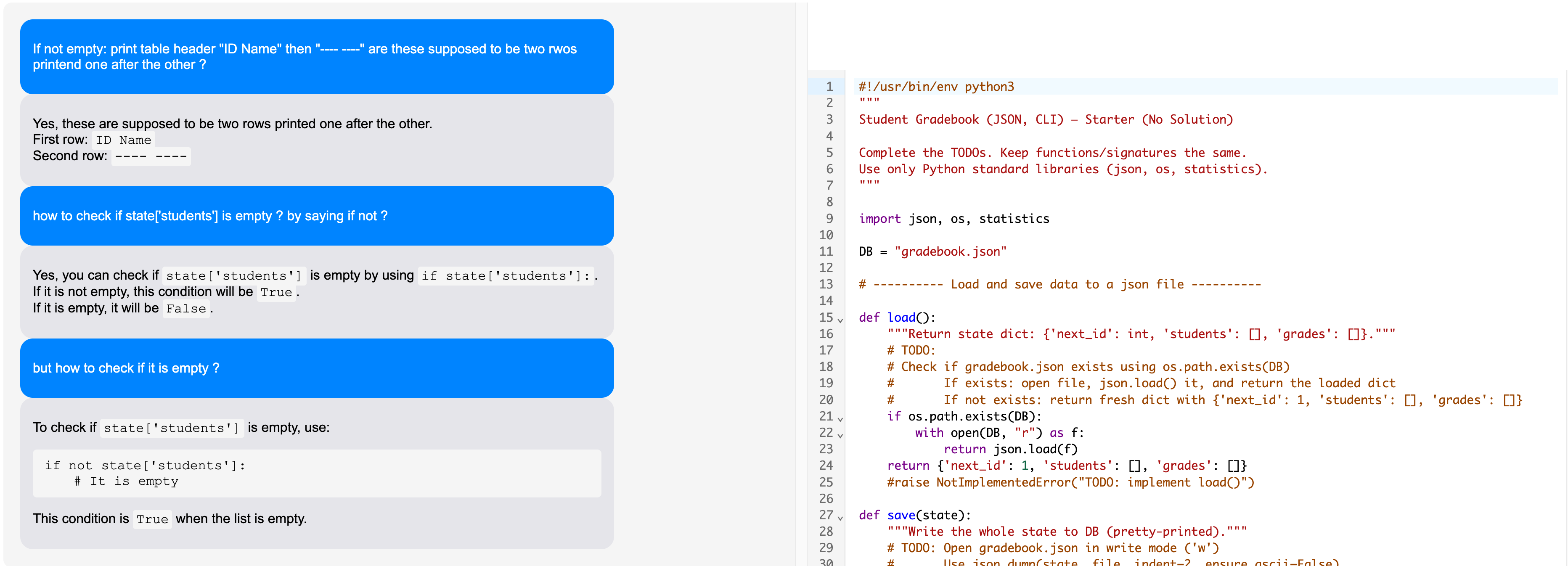}
    \caption{Baseline system for the user study. It represents conversation turns with AI agents (left) and code states (right).}
    \label{fig:baseline}
\end{figure*}
}

\newcommand{\figureComparisonSurvey}{
\begin{figure*}
    \centering
    \includegraphics[width=\linewidth]{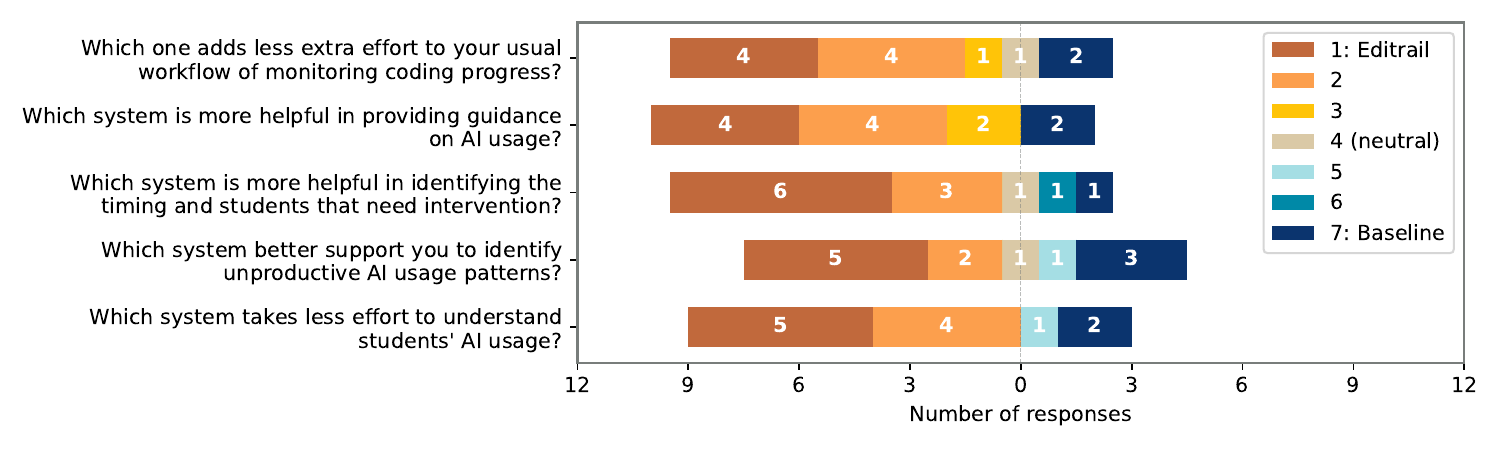}
    \caption{Results of the comparison survey. Participants answers questions (left) on a 7 likert-scale where 1 represents ``Prefer \sys{}'' and 7 represents ``Prefer Baseline''.}
    \label{fig:comparison-survey}
\end{figure*}
}

%% file: tables/tables.tex
\newcommand{\tableDataCollectionDemographics}{
\FloatBarrier{}
\begin{table}[ht]
\centering
\caption{Participant Demographics from the coding data collection study.}
\label{tab:data-collection-demographics}
\begin{tabular}{llll}
\toprule
\textbf{PID} & \textbf{Gender} & \textbf{Experience} & \textbf{Role} \\
\midrule
P1  & Man      & 7 years       & PhD Student \\
P2  & Nonbinary & 6 years       & Student \\
P3  & Woman    & 1--2 years    & Student \\
P4  & Man      & 2 years       & Student / Software Developer \\
P5  & Man      & 7 years       & PhD Student \\
P6  & Woman    & 2 years       & Student \\
P7  & Man      & 6 years       & Student \\
P8  & Woman    & 3 years       & Data Engineer \\
P9  & Man      & 3 years       & Student \\
P10 & Man      & $\sim$2 years & Technology Consultant \\
P11 & Man      & 1--2 years    & Student \\
P12 & Man      & $\sim$5 years & Data Analyst \\
P13 & Woman    & 15 months     & Student \\
P14 & Woman    & 4 years       & Developer \\
P15 & Woman    & $<$3 months   & UX Researcher \\
P16 & Man      & 9 years       & PhD Student \\
P17 & Woman    & 5 years       & Software Engineer \\
P18 & Man      & $\sim$1 year  & PhD Student \\
P19 & Man      & 8 years       & PhD Student \\
P20 & Woman    & 4 years       & Software Engineer \\
\bottomrule
\end{tabular}
\end{table}
\FloatBarrier{}
}

\newcommand{\tableUserStudyDemographics}{
\FloatBarrier{}
\begin{table}[ht]
\centering
\caption{Participant Demographics from the user study. We report their Python coding experience, courses they teach, and their professions.}
\label{tab:user-study-demographics}
\begin{tabular}{llp{6cm}l}  
\toprule
\textbf{ID} & \textbf{Python Experience (Years)} & \textbf{Course / Topic} & \textbf{Profession} \\
\midrule
1  & 10  & (1) Python – beginner to intermediate, focusing on data science and ML; (2) C for embedding systems, beginner & Teaching Assistant \\
2  & 6   & Beginner/Intermediate & Python Tutor \& Instructional Aide \\
3  & 7   & Intermediate & Teaching Assistant, Instructor \\
4  & 9   & None & Lead Virtual Engineer \\
5  & 6   & None & Graduate Student Researcher \\
6  & 9   & Use Python to implement basic database system including basic database operations & Teaching Assistant \\
7  & 8   & Python, beginner & Tutor \\
8  & 4   & Beginner Python & Teaching Assistant \\
9  & 4+  & Python, intermediate & Teaching Assistant \\
10 & 4   & Django & Graduate Student Instructor \\
11 & 4   & Models of Social Information Processing & Graduate Student Instructor \\
12 & 4   & Introductory Computer Science & Graduate Student Instructor \\
\bottomrule
\end{tabular}
\end{table}
\FloatBarrier{}
}

\newcommand{\tableSystemSurvey}{
\FloatBarrier{}
\begin{table}[t]
\centering
\caption{Participant's self-report results by conditions. Participants found them had better understanding on students' AI usage, providing guidance, and students that need intervention using \sys{} than the baseline.}
\label{tab:outcomes}
\setlength{\tabcolsep}{8pt} 
\renewcommand{\arraystretch}{1.3} 

\begin{tabular}{%
    p{0.25\linewidth}@{\hspace{1em}}                                
    l                                  
    r                                  
    r                                  
    >{\centering\arraybackslash}p{0.40\linewidth} 
}
\toprule
\textbf{Measures} & \textbf{Condition} & $M$ & $SD$ & 1 (little understanding) to 7 (good understanding) \\
\midrule

\multirow{2}{*}{\makecell[l]{Students' AI usage patterns}}
    & \sys{} & 5.92 & 0.90
    & \multirow{2}{*}{\includegraphics[height=1.3cm,keepaspectratio]{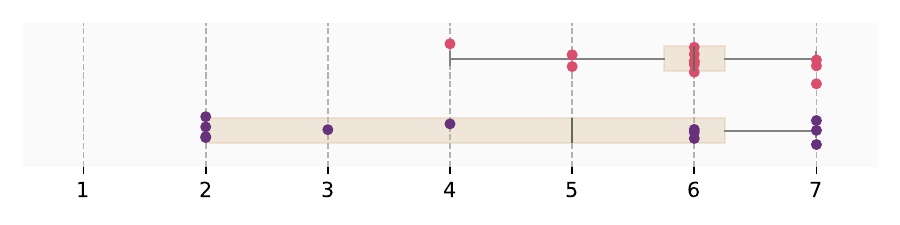}} \\
    & Baseline  & 4.50 & 2.20 & \\[2pt]
\addlinespace[4pt]

\multirow{2}{*}{\makecell[l]{Targeted, learning-oriented\\ guidance on AI usage}}
    & \sys{} & 5.67 &  1.50
    & \multirow{2}{*}{\includegraphics[height=1.3cm,keepaspectratio]{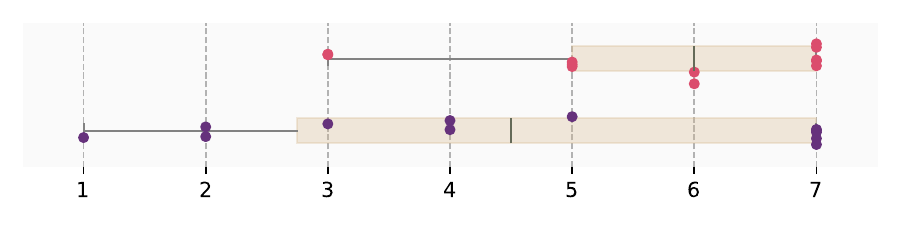}} \\
    & Baseline  & 4.67 & 2.31 & \\[2pt]
\addlinespace[4pt]

\multirow{2}{*}{\makecell[l]{Students that might need\\ intervention}}
    & \sys{} & 6.33 & 0.78
    & \multirow{2}{*}{\includegraphics[height=1.3cm,keepaspectratio]{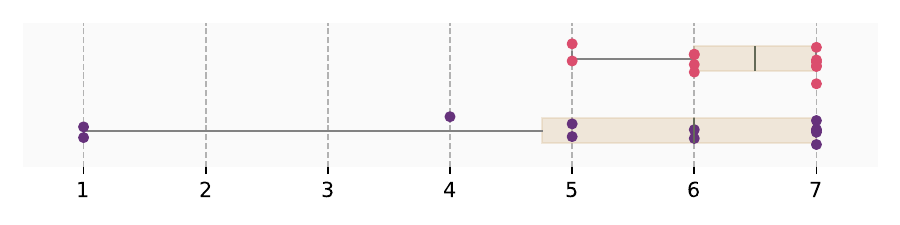}} \\
    & Baseline  & 5.25 & 2.22 & \\[2pt]
\addlinespace[4pt]

\bottomrule
\end{tabular}
\end{table}
\FloatBarrier{}
}

\newcommand{\tableBadAIUsage}{
\FloatBarrier{}
\begin{table}[ht]
\centering
\caption{Inappropriate AI Usage Patterns Reported by Instructors in the formative study. The table lists recurring behaviors observed by instructors. \replaced[id=az]{For each pattern, it lists how many participants mentioned it and a representative quote from participants.}{Counts indicate how many participants mentioned each issue.}}
\label{tab:bad-ai-usage}
\begin{tabular}{p{0.43\textwidth} c p{0.45\textwidth}}
\toprule
\textbf{Inappropriate AI Usage Patterns} & \textbf{\# participants} & \textbf{Quote} \\
\midrule
\replaced[id=az]{Uncritical use of AI-generated solutions, e.g., pasting without understanding, full AI-written submissions}{Students paste AI-generated code or answers without understanding or modifying them. Students also submit entire assignments written by AI without understanding the logic or algorithms involved.} & 15 &\added[id=az]{\myquote{Most frequently students copy homework problems into ChatGPT or another LLM, and paste the code results. We often see submissions where the students accidentally copied extra text from the LLM like the next question prompt, or left certain wording in the form the LLM uses.}} \\
\replaced[id=az]{AI use on assessments intended to measure individual understanding}{Students use AI even when assessments are meant to test individual understanding, e.g., programming assignments, paper reviews, etc.} & 7 & \added[id=az]{\myquote{The direct use of AI to trivially solve assigned problems that are explicitly intended to assess students' individual programming learning outcomes.}}\\
\replaced[id=az]{Code indicating external generation, e.g., advanced syntax, imported libraries}{Submissions contain advanced or deprecated code not taught in class, indicating outside (AI) generation.} & 3 & \added[id=az]{\myquote{When intro students submit solutions with complex syntax not covered in class, I assume they either used AI or a solution directly from Stack Overflow.}} \\
\replaced[id=az]{Undisclosed or policy-violating AI use}{Many students violate policies by using AI without disclosure or attribution, akin to academic dishonesty.} & 2 & \added[id=az]{\myquote{In the presence of my policies that allow students to use AI so long as they fully cite all their sources and prompts, students are still not being transparent about their use.}} \\
\replaced[id=az]{Shallow comprehension, e.g., lack fundamental skills}{Students don't learn much – they fail to build foundational skills.} & 2 & \added[id=az]{\myquote{[students] don't actually take any time to learn the material or understand the solution}}\\
Students struggle during office hours or interviews to explain AI-generated code, reflecting a lack of comprehension of concepts. & 2 & \added[id=az]{\myquote{Typically, student use of AI for coding is detected during office hours.  Such students have code that does not work and cannot express explain how their code works in debugging with an instructor. }}\\
\replaced[id=az]{Misinterpretation of Copilot usage}{Some students assume GitHub Copilot's suggestions are part of instructor-provided code.} & 2 & \added[id=az]{\myquote{ When the copilot add some code automatically, they sometimes mistaken that as part of the starter code provided by the instructors.}}\\
\replaced[id=az]{AI-generated artifacts showing conceptual misunderstandings, e.g., UI images, design docs}{AI-generated images used in UI or software design documentation often reflect misunderstandings of key concepts.} & 1 & \added[id=az]{\myquote{Some students/groups use AI-generated images in a document to describe a software design, and these designs are always inaccurate.}}\\
\replaced[id=az]{Identical AI-generated bugs across students}{Identical bugs from ChatGPT-generated code appear across multiple student submissions.} & 1 & \added[id=az]{\myquote{I observed a few students fully relying on GenAIs to write their coding homework and having the same bug as ChatGPT's solution. }}\\
\bottomrule
\end{tabular}
\end{table}
\FloatBarrier{}
}

%% file: sections/authors.tex
\author{Ashley Ge Zhang}
\orcid{0000-0001-5978-3714}
\affiliation{
    \institution{University of Michigan}
    \city{Ann Arbor}
    \state{Michigan}
    \country{USA}
}
\email{gezh@umich.edu}

\author{Yan-Ru Jhou}
\orcid{0009-0000-9987-0492}
\affiliation{
    \institution{University of Michigan}
    \city{Ann Arbor}
    \state{Michigan}
    \country{USA}
}
\email{yanruj@umich.edu}

\author{Yinuo Yang}
\orcid{0009-0003-8184-719X}
\affiliation{
    \institution{University of Notre Dame}
    \city{Notre Dame}
    \state{Indiana}
    \country{USA}
}
\email{yinooyang@nd.edu}

\author{Shamita Rao}
\orcid{0009-0001-7761-5615}
\affiliation{
    \institution{University of Michigan}
    \city{Ann Arbor}
    \state{Michigan}
    \country{USA}
}
\email{shamita@umich.edu}

\author{Maryam Arab}
\orcid{0000-0001-9040-4313}
\affiliation{
    \institution{University of Michigan}
    \city{Ann Arbor}
    \state{Michigan}
    \country{USA}
}
\email{maryarab@umich.edu}

\author{Yan Chen}
\orcid{0000-0002-1646-6935}
\affiliation{
    \institution{Virginia Tech}
    \city{Blacksburg}
    \state{Virginia}
    \country{USA}
}
\email{ych@vt.edu}

\author{Steve Oney}
\orcid{0000-0002-5823-1499}
\affiliation{
    \institution{University of Michigan}
    \city{Ann Arbor}
    \state{Michigan}
    \country{USA}
}
\email{soney@umich.edu}

%% file: sections/00-abstract.tex
\begin{abstract}\label{abstract}
Programming instructors have diverse philosophies about integrating generative AI into their classes.
Some encourage students to use AI, while others restrict or forbid it.
Regardless of their approach, all instructors benefit from understanding how their students actually use AI while writing code\added[id=az]{~\cite{sheard2024instructor, halaweh2023chatgpt, adnin2025examining, lau2023ban}}.
Such insight helps instructors assess whether AI use aligns with their pedagogical goals, enables timely intervention when they find unproductive usage patterns, and establishes effective policies for AI use.
However, our survey with programming instructors found that many instructors lack visibility into how students use AI in their code-writing processes.
To address this challenge, we introduce \sys{}, an interactive system that enables instructors to track students' AI usage, create personalized assessments, and provide timely interventions, all within the workflow of monitoring coding histories. 
We found that \sys{} enables instructors to detect AI use that conflicts with pedagogical goals accurately and to determine when and which students require intervention.
\end{abstract}

%% file: sections/document.tex
\input{sections/01-introduction}
\input{sections/02-relatedWork}

\input{sections/03-needFindingStudy}

\input{sections/04-systemDesign}

\input{sections/08-dataCollection}
\input{sections/05-userStudy}
\input{sections/06-discussion}
\input{sections/07-conclusion}

%% file: sections/01-introduction.tex
\section{INTRODUCTION}\label{sec:intro}

Recent advances in large language models (LLMs) present both opportunities and challenges in computer science education (CSEd). While LLMs can boost students’ productivity~\cite{shihab2025effects} and provide flexible feedback~\cite{phung2024automating}, they also raise concerns about over-reliance and reduced engagement in the learning process. Prior work highlights unproductive AI use, such as the illusion of competence (completing tasks without understanding the code)~\cite{prather2024widening}, as well as repeated re-prompting instead of debugging and blindly accepting incorrect outputs~\cite{shihab2025effects, ferdowsi2024validating}.

Programming instructors hold different philosophies on how to best integrate AI into their courses. Some encourage its use to prepare students for future practice, while others restrict or prohibit it to prevent dependency. Regardless of approach, instructors benefit from understanding how students use AI~\cite{chen2024stugptviz}. For instance, students who bypass learning may show patterns of frequent AI-generated edits, whereas those who engage critically often make iterative edits to verify and adapt AI suggestions\added[id=az]{~\cite{kazemitabaar2023novices, prather2024widening, guner2025ai, rahe2025programming}}.
\added[id=az]{The nuances in students’ editing patterns may reflect different forms of engagement across contexts. This makes it essential to surface students’ interaction processes, enabling instructors to interpret these behaviors in ways that align with their own pedagogical goals.}

\replaced[id=az]{In this paper, we report findings from a formative study with programming instructors that,}{However, our formative study with programming instructors revealed that,} \added[id=az]{instead of determining which behaviors are productive or unproductive in the abstract, the main challenge is that} they lack visibility into students’ AI usage patterns and need tools to provide targeted, personalized guidance that supports learning. Without such support, instructors often rely on indirect signals, such as unusual code style, advanced syntax not covered in class, or overly polished solutions. These indirect signals can be inconsistent and unreliable. This lack of transparency makes it difficult to distinguish between productive AI use that supports learning and harmful patterns that bypass understanding, limiting instructors’ ability to intervene at the right time, align AI use with pedagogical goals, and \replaced[id=az]{make contextualized judgments}{develop nuanced classroom policies}. Prior work has sought to address this challenge by surfacing students’ interactions with AI systems, such as visualizing AI-student dialogues~\cite{chen2024stugptviz}, but dialogue alone is insufficient to capture how students actually use and adapt AI-generated content in their learning.

To address this challenge, we propose \sys{}, a system for understanding student–AI interactions and facilitating a feedback loop among instructors, students, and AI. \sys{} integrates AI usage monitoring into existing code-progress tracking workflows by visualizing AI contributions as trails in students’ edit histories. With these visualizations, instructors can \replaced[id=az]{use fine-grained contextual evidence to diagnose misunderstandings in AI-usage}{more easily identify unproductive AI usage patterns}, deliver in-situ assessments, and provide timely support. For example, instructors can generate targeted quiz questions on concepts where students relied heavily on AI.

To evaluate the efficacy of \sys{} in understanding students’ AI usage and supporting timely intervention, we collected coding histories and AI interactions from 20 students across two Python programming problems and conducted a within-subject study with 12 instructors. Our findings show that by visualizing AI code contributions as trails in students’ edits, \sys{} enables instructors to identify AI usage \replaced[id=az]{more accurately and efficiently than a baseline system that shows only code and chat logs.}{with twice the accuracy of a baseline system and in less time}. \sys{} also reveals new opportunities for designing personalized guidance around students’ AI use. By bridging the visibility gap between students’ interactions with AI and what instructors can observe, \sys{} helps instructors align interventions with pedagogical goals and inform effective classroom AI policies. This paper makes three contributions:

\begin{enumerate}
    \item A need finding survey of 27 lead programming instructors in 15 universities, spanning multiple continents and course levels, identifies the limited visibility of instructors into student-AI interactions, the resulting challenges in enforcing AI usage policies, and design goals for tools to support monitoring and guidance.
    \item A dataset of 20 students' coding histories on two Python programming problems, capturing 188--319 lines of code, up to 2610 edits, and 42 AI chats. 
    \item An interactive tool (\sys{}) that enables instructors to identify AI usage patterns and deliver in-situ interventions to students. 
    \item Evidence from a comparison study showing that \sys{} bridges the visibility gap between students’ AI use and what instructors can observe, allowing instructors to align AI policies and interventions with their pedagogical goals. 
\end{enumerate}

%% file: sections/02-relatedWork.tex
\section{RELATED WORK}\label{sec:rw}

\subsection{AI in Programming Education}\label{sec:rw-ai}

Generative AI has transformed programming education by automating tasks such as diagnosing bugs, generating feedback, producing code examples, and creating learning materials~\cite{balse2023evaluating, tang2024sphere, hou2024codetailor, kazemitabaar2024exploring}. AI assistants like Codex and CodeHelp have shown promise in helping novices by providing hints, scaffolding, and on-demand support~\cite{finnie2022robots, denny2024desirable}. Studies also highlight cognitive engagement techniques—such as step-by-step prompting or worked examples—that can help learners integrate AI suggestions productively~\cite{kazemitabaar2024exploring, jury2024evaluating}.

At the same time, risks accompany these benefits. AI-generated content is often inaccurate or misleading~\cite{becker2023programming}, and over-reliance may lead novices to copy–paste solutions, skip validation, or shepherd prompts toward full answers without developing problem-solving skills~\cite{prather2023s, amoozadeh2024student}. Classroom deployments report both benefits and harms, underscoring the tension between efficiency and deep learning~\cite{prather2024widening}.
\added[id=az]{In a deployed study with 69 novices, Kazemitabaar et al. found that learners with more prior programming competency may benefit more from AI code generators, introducing design needs of control on over-reliance and support writing prompts, which requires support on understanding students' AI usage~\cite{kazemitabaar2023studying}.}
\added[id=az]{While AI offers powerful opportunities, it also poses risks: suggestions can be inaccurate~\cite{becker2023programming}, and direct solution generation may encourage copy-paste behaviors that undermine learning~\cite{kazemitabaar2024codeaid}. Preparing both students and instructors to use AI responsibly is therefore critical~\cite{pedro2019artificial}.}

\added[id=az]{For students to effectively learn programming with GenAI, researchers have identified practical strategies, including breaking problems down, iteratively refining prompts, and validating AI-generated code~\cite{porter2024learn}. Effective AI use requires substantial meta-cognitive activity, including planning, monitoring, and critically evaluating AI-generated content~\cite{tankelevitch2024metacognitive}.}
To mitigate these risks, researchers have proposed designs such as teachable agents that students can debug~\cite{ma2024teach}, live programming tools that surface runtime behavior to validate AI code~\cite{ferdowsi2024validating}, and interfaces that promote trust, transparency, and control~\cite{kazemitabaar2024codeaid, pozdniakov2024large}. Yet, there remains a need for tools that help instructors monitor student–AI interactions, interpret whether usage aligns with pedagogical goals, and provide timely guidance to foster productive habits.

\subsection{Understanding Student-AI Interaction in Programming}

A growing body of work investigates how students engage with generative AI tools and the challenges these interactions introduce. Case studies in introductory programming (CS1) have shown that students sometimes over-rely on AI, for example, by pasting entire task descriptions into ChatGPT without contributing their own effort, or by neglecting to verify AI-generated solutions even when they recognize errors~\cite{amoozadeh2024student}. Similarly, Prather et al. found that students often blindly accept AI suggestions, reinforcing misconceptions and encouraging surface-level completion rather than deeper problem solving~\cite{prather2023s}.

To address these issues, researchers have proposed strategies to encourage more productive engagement. For example, Kazemitabaar et al. deployed CodeAid, an LLM-powered assistant in a semester-long course, and distilled design principles for classroom deployment: exploiting AI’s unique advantages, balancing the directness of responses, and supporting trust and transparency~\cite{kazemitabaar2024codeaid}. In follow-up work, they examined cognitive engagement techniques, showing that step-by-step problem-solving guidance with interactive prompts before revealing a final solution was particularly effective at fostering learning~\cite{kazemitabaar2024exploring}.

In addition to fostering good usage on the students' side, complementary approaches have explored how to make student–AI interactions more visible and interpretable. For instance, Chen et al. introduced StuGPTViz, a visual analytics system that highlights temporal patterns in student use of ChatGPT~\cite{chen2024stugptviz}. Beyond education, work in HCI has proposed methods for interpreting LLM outputs in decision-making contexts~\cite{kahng2024llm}, suggesting opportunities to adapt similar techniques for programming pedagogy.

However, these works have largely focused on AI outputs, such as what was prompted and what the model generated, without capturing how students engaged with these suggestions, whether they understood them, or how effectively they integrated them into their code. Our work fills this gap by integrating AI outcomes with the processes of adoption and adaptation, giving instructors visibility into how students interpret, modify, or struggle with AI-generated code. 
This process-oriented view helps instructors distinguish productive from unproductive patterns and provide targeted guidance aligned with pedagogical goals.

\subsection{Providing Feedback based on Students' Progress}\label{sec:rw-progress}


Prior work in education and learning sciences has highlighted the important role of feedback in sustaining student engagement and persistence. Studies on automatically generated feedback for assignments have compared different strategies and shown how variation in feedback design can influence learning outcomes~\cite{van2024investigation}. A consistent finding across this literature is that students benefit more from active engagement in problem-solving activities than from passive content consumption~\cite{koedinger2015learning, koedinger2016doer, van2023doer}. Within this context, feedback is understood not only as corrective but also as motivational, supporting students in persisting after errors and refining their understanding~\cite{vanlehn2011relative}.

Building on this foundation, researchers have investigated scalable approaches to collecting and analyzing student feedback and emotions. Sentiment analysis, for example, has been shown to efficiently capture learners’ affective states, with the potential to mitigate anonymity and feedback gaps in online education~\cite{neumann2021capturing}. Related work analyzing sentiment in social media~\cite{permana2017naive}, longitudinal course evaluations~\cite{camacho2018longitudinal}, and even small corpora such as 105 learning diaries~\cite{munezero2013exploiting} demonstrates that meaningful affective signals can be extracted even from limited datasets. These studies point to the possibility that effective feedback systems may not need to rely solely on large-scale data; instead, they can leverage methods that extract actionable insights from sparse information to support students’ cognitive and emotional needs at scale.

Specifically in the context of programming education, Jeuring et al. synthesized guidelines for when and how to provide formative feedback and hints to novice programmers~\cite{jeuring2022towards}. Their recommendations emphasize that feedback should be timely and closely tied to students’ coding actions, striking a balance between being directive enough to guide progress and open enough to foster problem-solving autonomy. They also stress the importance of adapting feedback to different stages of the learning process—such as error detection, strategy refinement, and conceptual understanding—and propose scaffolding strategies that gradually increase in explicitness, enabling students to attempt self-correction before receiving more detailed guidance. Research has shown that feedback alone is often insufficient for learners who are stuck; additional hints are necessary to suggest concrete next steps and help learners re-engage with the problem ~\cite {cosejo2009towards, marwan2021investigating,vanlehn2006behavior}. 

\added[id=az]{Prior work has introduced tools to help instructors understand students’ code in order to provide targeted and timely feedback~\cite{zhang2023vizprog, glassman2015overcode, guo2015codeopticon, head2017writing, zhang2024cflow}. Some systems cluster code submissions to reveal common approaches and mistakes, enabling reusable feedback at scale~\cite{glassman2015overcode, zhang2024cflow, head2017writing}. However, these techniques primarily operate on final submissions and overlook the dynamic process of how code is written. CodeOpticon addresses process-level visibility by streaming multiple students’ live editors~\cite{guo2015codeopticon}, but this requires intensive manual monitoring and does not scale to large classrooms. VizProg reduces this burden by visualizing coding progress as dynamic points on a 2D map~\cite{zhang2023vizprog}, yet these tools focus on progress rather than the provenance of code.
More recently, tools such as Meta-Manager visualize metadata about code origins, for example labeling activities like copy–paste or Stack Overflow search~\cite{horvath2024meta}. However, such high-level summaries provide only coarse signals and are insufficient for instructors to fully understand how students use AI while programming. \sys{} fills this gap by directly visualizing contributions from both students and AI within coding histories, enabling instructors to identify AI usage patterns and deliver timely, targeted support.}

%% file: sections/03-needFindingStudy.tex
\section{NEED FINDING STUDY}
To better understand instructors' views on effective and inappropriate AI usage, we conducted a needs-finding study using surveys.
The survey provides a broad overview of instructors' perspectives based on their observations in programming courses. 
We used thematic analysis to qualitatively code the survey responses, which inform the system design to support understanding and intervening in student-AI interactions.

\subsection{Study Method}
Prior work has explored the opportunities and challenges of AI tools in programming education on a small scale~\cite{prather2024widening,becker2023programming}. 
These studies focused on specific course types, regions, or AI tools. 
Our survey aims to extend the findings to a diverse, global audience of programming instructors at all levels.
We aim to gather instructors' views on students' AI usage in programming courses and the rationale behind their policies on AI usage in class.
Our survey includes questions on effective and inappropriate AI use, methods for detecting misuse, institutional AI policies, and the rationale behind these policies.
We also collected demographic data such as teaching experience, institution, and course levels.

\subsubsection{Participants}\label{sec:formative-participants}
Our target population is university-level programming instructors teaching `traditional' (in-person, synchronous) courses ranging from introductory to advanced topics. We include a wide array of programming courses (e.g., Python programming, web development, and machine learning) to capture diverse experiences with AI integration in the curriculum.

Rather than distributing a broad public poll, we adopted a targeted recruitment strategy where we reached out to instructors directly.
Open calls often yield self-selected responses concentrated in a few regions, and critically, they provide no reliable way to verify whether respondents are, in fact, programming instructors.
We also specifically sought \emph{lead instructors}, since they are typically the ones who set course-level policies on AI usage and shape how such tools are integrated into programming curricula.
To recruit participants, we followed these steps:

\begin{itemize}

    \item \textbf{Identify universities}.
    We started by identifying multiple institutions to recruit from. 
    We focused recruitment on (1) the `top' 50 universities worldwide, using an established ranking (Times Higher Education)\footnote{https://www.timeshighereducation.com} and (2) the three institutions that the authors of this paper are affiliated with.
    Our choice to focus on `highly-ranked' institutions was not because these universities are inherently more valuable, but because the list provided a tractable way to select institutions that span multiple continents, maintain large programming curricula across course levels, and have the resources and early exposure to AI tools needed to support relevant classroom practices. 
    This approach ensured both diversity and feasibility in recruitment, while avoiding the impracticality of attempting to reach the full global population of programming instructors.
    \added[id=az]{Of the 50 universities, we proceeded with 41, primarily because some lacked publicly available instructor information.}
    
    \item \textbf{Identify relevant courses and instructors}. For each selected university, we identified a list of programming-related courses for the current academic year.
    \added[id=az]{We focused on ``high-resource'' programming languages and concepts (i.e., topics where AI may be effective). This included courses in Python, C, algorithms, and data structures, spanning both introductory and advanced courses. We prioritized courses with substantial coding components, a high student-instructor ratio, and assignments that AI tools are more likely to be capable of completing. }
    Both undergraduate- and graduate-level courses were included.
    We excluded seminars, capstone projects, and theory-based courses without substantial coding components. 
    Short-term workshops or bootcamps outside the standard curriculum were also excluded.
    We then extracted the names and contact information of the lead instructors.
    \item \textbf{Email outreach}. We sent personalized emails to each instructor, including: (1) a description of the study’s goals and relevance, (2) a link to the online survey, and (3) a request for referrals, encouraging instructors to forward the invitation or suggest potential participants.
\end{itemize}
To encourage participation, participants would be entered into a lottery for a \$10 (USD) Amazon gift card and receive access to a dataset of anonymized survey responses.
Of the 208 instructors we contacted, 27 of them responded to the survey. 
The participants were from 15 universities. 
The respondents represent a range of teaching experience, with the majority having between 3 and 15 years in instructional roles. Three respondents reported less than three years of experience, and others reported more than 15 years, including 4 participants with over 25 years.
In terms of instructional role, all of the respondents identified as Head or Lead Instructors, indicating that the perspectives collected largely reflect those who hold primary responsibility for course design, instruction, and grading. 
Regarding course levels taught, the data show that most respondents are involved in introductory (21) and intermediate (23) courses, which constitute the foundation of many programming curricula. Some participants also teach advanced (11) and applied/specialized (9) courses. A small number of respondents reported teaching across all four levels, reflecting a broad instructional scope.

\subsubsection{Data analysis}
\replaced[id=az]{
We conducted an inductive thematic analysis~\cite{braun2006using}, beginning with open coding~\cite{strauss1994grounded} of all responses by one author to generate initial codes. 
A second author then coded the dataset using this preliminary thematic structure, refining and expanding the codes as needed.
The two authors met regularly to compare interpretations, discuss discrepancies, and iteratively refine the codebook until a shared understanding was reached.
Six authors reviewed the emerging themes and finalized the thematic structure. 
In parallel, we conducted a descriptive quantitative analysis of closed-ended and countable responses. For items where participants mentioned behaviors, policies, or detection strategies, we computed simple frequency counts (e.g., how many instructors reported a pattern) to contextualize the qualitative findings. 
The final themes covered: (\textit{i}) good and bad AI usage in programming courses, (\textit{ii}) policies and rationales regarding AI usage, and (\textit{iii}) ways of detecting AI usage and instructors' thoughts on tool incorporation.}{Two authors conducted an initial qualitative analysis via open coding~\cite{strauss1994grounded} to identify common themes related to the research questions. The themes were discussed, refined, and categorized by six authors. }
More specifically, we categorized \replaced[id=az]{the key findings}{the themes} into three aspects: (\textit{i}) the need of better understanding of students' AI usage, (\textit{ii}) the difficulty of tracking AI usage, and (\textit{iii}) unclear AI usage policies in class. Based on our findings, we derived three design goals for understanding and guiding student-AI interaction in programming courses.

\subsection{Instructors Lack a Clear Understanding of Students' AI Usage. }
\tableBadAIUsage{}

We found that instructors generally lack a clear understanding of students’ AI usage. When asked about helpful or harmful usage patterns, their responses were often based on isolated behaviors from individual students, lacking broader generalizability. We synthesized 9 high-level patterns of inappropriate AI usage reported by instructors (Table~\ref{tab:bad-ai-usage}). \added[id=az]{The most common issues involved students submitting AI-generated code without understanding it, using AI on assessments intended to measure individual learning, or producing submissions that were unusually advanced or inconsistent with course content, indicating external generation. For each pattern, we include a representative quote to illustrate how instructors observed and interpreted these behaviors.}

Currently, instructors have limited tool support to detect or prevent inappropriate AI usage in programming courses. Twelve participants reported relying on manual checks—such as reviewing code quality, advanced syntax, or unusual references—as indirect indicators of misuse. 
\added[id=az]{For example, participants mentioned that \myquote{AI-generated codes are often too beautiful and concise than students' codes} and they \myquote{look into code submitted by students manually and we have code reviews for those who achieve a grade that is 80\% or higher.} Additionally, \myquote{importing external libraries and using coding conventions outside the scope of the course can be indicators.}}
Three reported verbally checking in with students or asking them to walk through their code to assess their understanding. 
\added[id=az]{\mylongquote{We have our teaching assistants do small verbal check-ins with the student to ask them questions about their solution. These questions are not meant to take very long, but anyone who has genuinely done their own assignment will be able to answer them very easily.}}
One instructor required students to \replaced[id=az]{\myquote{upload chat logs and explain how they use AI as part of their code documentation and report}}{upload chat logs and explain their AI usage along with their assignments}. 
Three others mentioned using tools like Turnitin\footnote{https://www.turnitin.com}, MOSS\footnote{https://theory.stanford.edu/~aiken/moss}, or Gradescope\footnote{https://www.gradescope.com} to detect plagiarism and analyze document. 
However, these tools either capture only specific patterns or focus solely on the final submission, missing the student’s problem-solving process.
\added[id=az]{\mylongquote{In the past, the courses I've taught have also used online IDEs that provide keystroke history logs to instructors: these have made it very obvious when students paste in code, especially when that code includes chat context from an LLM. I do not use AI detector platforms.}}

Our findings suggest that instructors currently rely on ad hoc, experience-driven heuristics. We found no consistent or scalable rules for identifying inappropriate AI use that could be feasibly implemented across classrooms.
\added[id=az]{\mylongquote{It is effectively impossible to prevent students from using it on homeworks, or to accurately detect and penalize it in the context of programming.}}
Ultimately, instructors care most about whether students skip the learning process when using AI. But without direct evidence, they must infer student understanding solely from the final submission. 
\replaced[id=az]{One participant brought up a potential risk:\mylongquote{They may get marks/pass the autograder, but fundamentally they rarely seem to be able to sort out the issues that are confusing them in the first place. This eventually snowballs into even bigger problems down the line.}}{This becomes risky when students produce seemingly correct code without addressing underlying misconceptions, allowing small issues to snowball into deeper misunderstandings over time.}

\subsection{Tracking Students' AI Usage Is Difficult Due to Process Invisibility and Behavioral Variation}\label{sec:formative-variation}

Instructors report several factors that make it challenging for them to track how students use AI. Most courses only give them access to students’ final submissions, offering little visibility into the process the students followed. This makes it difficult to see how students prompt LLMs, adapt generated code, or how much work is authored by themselves versus the AI. Instructors instead rely on indirect signals in the final code or on students’ self-reports, both of which are unreliable.
\added[id=az]{\mylongquote{It is impossible to tell whether students used AI or not. We can't have any concrete evidence unless students confess.}}
Students also vary widely in how they use AI. Some rely heavily on AI-generated code, others edit extensively, and some use AI with restrictions. Current tools do not support detecting or categorizing these behaviors, leaving instructors to conduct detailed process reviews, which can be time-consuming and often impractical. As one participant noted, keystroke logs and chat transcripts might reveal usage patterns, but they are too complex for instructors and TAs with limited time to analyze.
\added[id=az]{\mylongquote{We manually look through all the submissions for the AI red flags we've found, and bring in those students to see if they actually understand their code. This is a lot of work, but because we don't have too many programming assignments per semester, we do it.}}

\added[id=az]{To address the challenge of lacking visibility into how students use AI and currently relying on indirect, unreliable signals, we proposed \dgbadge[teal]{DG1}: provide visibility into students' AI usage.}

\subsection{Unclear Tracking of AI Usage Leads to Inconsistent and Unenforceable AI policies}\label{sec:formative-policy}

These challenges in tracking AI usage directly affect how instructors create course policies. Policies varied by personal philosophy and course level. In introductory courses (e.g., CS1 or CS2), 8 instructors prohibited AI use entirely, while 12 others allowed it under conditional restrictions; no one reported fully open use. Intermediate courses largely allowed limited use of AI tools, often paired with discussions of LLMs, with only one instructor having a strict prohibition. Advanced courses did not show prohibitions, instead integrating AI as a tool with disclosure requirements. Applied and specialized courses were the most open, generally allowing AI except in specific assessments. In general, the trend shifts from stricter control in introductory courses to greater flexibility and integration in advanced and applied contexts.

Despite these differences, most instructors shared a common rationale behind their policies: ensuring AI use does not hurt students' ability to learn. They emphasized the need for students to first develop independent problem-solving skills before turning to AI:

\mylongquote{I cannot force students not to use AI. I would like them to be mindful of the use that they make of it, and I try to stress the fact that they need to understand how to do something themselves before relying on the help of the AI, which will rob them of the processes needed to deepen their understanding.}

At the same time, our findings suggest that the binary rules `allow' or `ban' are insufficient. Without ways to reliably identify inappropriate usage, policies were difficult to enforce and sometimes created unfair situations where students who followed restrictions were disadvantaged compared to peers who ignored them:

\mylongquote{I realized limiting or prohibiting the use of AI is quite unenforceable, and we end up in an unfair situation where some students respect the rule and end up at a disadvantage because others still use AI, and we can't really tell.}

Even when policies explicitly required citation or disclosure, instructors noted that students were not always transparent:

\mylongquote{In the presence of my policies that allow students to use AI so long as they fully cite all their sources and prompts, students are still not being transparent about their use.}

\added[id=az]{Difficult-to-enforce AI policies make it unclear how to respond to suspect inappropriate use. Without knowing the students’ process, it is difficult to determine whether a submission reflects misunderstanding, over-reliance on AI, or a genuine attempt to follow course expectations. This motivates \dgbadge[violet]{DG2}: enabling targeted, learning-oriented guidance on AI usage, so instructors can address misconceptions and support students productively rather than relying on punitive or policy-driven responses. }

\subsection{Design Goals to Support Monitoring and Guiding AI Usages}\label{sec:design-goals}

As some instructors pointed out in the study, AI usage in programming is unavoidable and should be integrated into teaching so that students use it in a helpful way. 
To help instructors effectively spot inappropriate AI usage patterns and provide personalized guidance, we concluded the following design goals (DGs) from the need-finding survey study:

\begin{itemize}
    \item \dgbadge[teal]{DG1}\label{dg:visibility} \textbf{Provide visibility into students' AI usage.} Students’ AI usage should be transparent and interpretable to instructors. The system could reveal key aspects of student-AI interaction, such as prompting, adaptation, and authorship balance. This reduces reliance on manual inspection or self-report, giving instructors the awareness they need to understand coding behaviors and intervene when necessary.
    
    \item \dgbadge[violet]{DG2}\label{dg:guidance} \textbf{Enable targeted, learning-oriented guidance on AI usage.} Visibility alone is not sufficient; instructors also need ways to act on this information. Systems should help instructors connect observed AI usage patterns with context-specific feedback that ensures students learn core concepts rather than bypass them. For example, when instructors see over-reliance on AI suggestions, the system could help them create feedback that addresses misconceptions, scaffolds problem-solving, and encourages productive use of AI. 
    
\end{itemize}

%% file: sections/04-systemDesign.tex
\section{System Design}
\figureSys{}

Guided by our design goals (Section~\ref{sec:design-goals}), we developed \sys{} to help instructors understand students' AI usage by visualizing student-AI interaction in code. 
\sys{} achieves this goal in several ways.
First, it allows instructors to monitor AI usage in real-time and identify unproductive usage patterns connected with students' code edits. 
Second, it enables instructors to create targeted, learning-oriented questions in situ to assess students' understanding. 
We describe the design of each of these facets of \sys{} in more detail below.

\subsection{\added[id=az]{\sys{}'s UI Overview}}\label{sec:ui-overview}

\added[id=az]{The interface of \sys{} consists of two coordinated views: a Timeline View (Figure \ref{fig:sys}.1) and a Code Content View (Figure \ref{fig:sys}.2). Together, they provide an overview of a student’s editing process and AI involvement.
The Timeline View summarizes the student’s keystroke and AI usage activity over time.
The x-axis encodes time from the start to the end of the coding session (Figure \ref{fig:sys}.a), while the y-axis represents code line numbers, with the top corresponding to the first line and the bottom to the last (Figure \ref{fig:sys}.b).
The light-gray background below the timeline represents the evolving length of the file: it expands or contracts as lines are inserted or deleted, helping instructors track structural changes.
Each edit appears as a colored marker positioned by time (x) and affected line (y) (Figure \ref{fig:sys}.c).
Blue markers indicate insertion edits and orange markers indicate deletion edits.
AI-involved edits are shown as colored overlay on the timeline (Figure \ref{fig:sys}.d): red for copy–paste, green for autocomplete, and pink for student-typed code that closely resembles AI output.
The Code Content View shows the actual program text aligned with the timeline (Figure \ref{fig:sys}.2), enabling instructors to inspect exactly how the code changed at any moment. Clicking a region of the timeline highlights the corresponding lines in the code view.
For longer files, where the timeline must be compressed vertically or the code viewer shows only a subset of lines, \sys{} provides a projection indicator that shows which portion of the code is currently visible (Figure \ref{fig:sys}.c). This helps instructors maintain orientation between the two views.
These two views work together to reveal students’ editing behavior and AI usage patterns. Building on this foundation, the following sections describe how \sys{}’s interaction techniques support our design goals.}

\subsection{Monitoring AI Usage in Context}

\replaced[id=az]{\sys{} supports instructors in monitoring AI usage by combining keystroke-level logs, AI interactions, and visual overlays that connect code provenance with code evolution. Below we describe the key interactions and how they support our design goals.}{\sys{} visualize each student's AI usage in their code edits through a timeline visualization (Figure~\ref{fig:sys}.1), along with their code content shown in a code editor (Figure~\ref{fig:sys}.2). }


\subsubsection{Tracking AI Usage When Students Write Code \dgbadge[teal]{DG1}.}
\replaced[id=az]{To increase visibility into AI usage, \sys{} first captures students’ interactions with AI coding tools and their keystroke-level edits through two lightweight mechanisms:}
{\sys{} includes several mechanisms for tracking students' AI usage.}
(1)~a VS Code extension records keystroke-level edit logs and interactions with AI coding assistants, and (2)~a browser extension monitors students' use of ChatGPT. Both would forward relevant interaction data \added[id=az]{to the instructor side}. 
These mechanisms are not intended to be foolproof---motivated and technically knowledgeable students could sidestep detection---but they provide instructors with a practical level of visibility that is otherwise unavailable.
\added[id=az]{The data collected through these mechanisms supported the following interactions.}

\subsubsection{Viewing Code Edits in Real-Time \dgbadge[teal]{DG1}.}
\figureUIDetailsTimeline{}
\replaced[id=az]{\sys{} adapts and extends prior work of a timeline-based visualization of keystroke edits}{We created a timeline visualization that extends prior work}~\cite{anonymous2025}.
The timeline visualization shows students' keystroke level edits and how their code evolves overtime \added[id=az]{using the keystroke-level data captured by our VS Code extension}. 
\replaced[id=az]{As described in Section~\ref{sec:ui-overview}, the timeline view shows an overview of code changes, mapping each edit to its time (x-axis) and location in the file (y-axis) (Figure~\ref{fig:sys}.a–b). Insertions and deletions appear as blue and orange markers, respectively, positioned according to when and where they occurred (Figure~\ref{fig:ui-details-timeline}.a).}{On the timeline, the X-axis represents time, where the left side represents when students start coding, and the right side represents their current code state or the final code state if the student has completed coding (Figure~\ref{fig:sys}.a). The Y-axis represents code lines, from top to bottom is from the first line of code of the students to their last line of code (Figure~\ref{fig:sys}.b).}

\sys{} \replaced[id=az]{represents}{extends this timeline in several ways (described in the subsubsections below) first, by representing} different code sources using shading.
The gray shading of the timeline represents\added[id=az]{ code file length, showing} how the students' code line grows and reduces over time (Figure~\ref{fig:sys}.\replaced[id=az]{1}{c}).
\added[id=az]{When the code file gets long and the edits get more frequent, the timeline could shrink. Users can zoom in on the timeline to view the actual code content of the changes on the timeline (Figure~\ref{fig:ui-details-timeline}.b, more details about how the code content is displayed could be found in our prior work~\cite{anonymous2025}). Users can also zoom in on the timeline to view the content that has been changed on the timeline (Figure~\ref{fig:ui-details-timeline}.b). AI-involved edits are shown as colored overlay on the timeline (Figure~\ref{fig:sys}.d, more details described in the following parts).}

\deleted[id=az]{The timeline visualizes students' keystroke-level edits through the change indicator in blue and orange (Figure~\ref{fig:ui-details}.a). The blue change indicator means adding characters to the code, while the orange ones indicate deletion. The position of the change indicators represents when and to which code line the edit happened. }

\subsubsection{Connecting AI Usage with Code Evolution \dgbadge[teal]{DG1} \dgbadge[violet]{DG2}}
\figureUIDetailsAI{}
Unlike systems that only present raw chat logs or static AI responses, \sys{} embeds AI usage directly within the temporal flow of edits (Figure~\ref{fig:sys}.d) \added[id=az]{to help instructor identify different AI usage patterns and provide targeted help}. 
\added[id=az]{On the timeline, \sys{} uses colored bars to represent messages sent from the student and AI, where blue represent student and gray represent AI (Figure~\ref{fig:ui-details-ai}.a). The bar's position on the x-axis represent when the message was sent, and the height represent message's word count. When hovering on the bar, users can see the actual message content.}
When a student adopts an AI suggestion, instructors can see exactly when it entered the code (Figure~\ref{fig:ui-details-ai}.c), how it evolved, and whether students refined, deleted, or repeatedly reverted it (Figure~\ref{fig:ui-details-ai}.d). \added[id=az]{The code lines edited are shown as colored overlays (Figure~\ref{fig:ui-details-ai}.c-d).}
Users can also locate the changes in code \added[id=az]{content view} by clicking the change indicators, the code \replaced[id=az]{content view}{editor} on the right would jump to that line of code (Figure~\ref{fig:ui-details-ai}.b).
\deleted[id=az]{Users can also zoom in on the timeline to view the content that has been changed on the timeline (Figure~\ref{fig:ui-details-timeline}.b). }
This contextualization reveals whether students meaningfully integrated AI code or relied on it uncritically. For example, patterns such as rapid copy–paste followed by minimal adaptation, or cycles of deletion and reinsertion, highlight potential misuse that would otherwise be hidden in the final submission.

\subsubsection{Differentiating AI-Generated vs. Student-Written Code \dgbadge[teal]{DG1} \dgbadge[violet]{DG2}}
\figureAISources{}
\sys{} explicitly marks code that originates from AI versus code authored or modified by students. 
\sys{} categorizes code originates from AI into three sources: (1) code that is copy-pasted from AI suggestions (Figure~\ref{fig:ai-sources}.a), (2) code that is directly inserted to editor by pressing `tab` to accept auto-completion suggestions from tools like Github Copilot (Figure~\ref{fig:ai-sources}.b), and (3) code that is manually typed by students but very similar to AI generated content as detected by the VSCode or browser extension (Figure~\ref{fig:ai-sources}.c). 
\sys{} mark three types of code that originate from AI with distinct color highlights (Figure~\ref{fig:sys}.d). Code that is copied and pasted is highlighted in red, autocomplete in green, and typed but similar to AI is in light pink. 
For any code that originates from AI, \sys{} visualize it as trails in code edit, making it clear to instructors which part is authored by students themselves vs. contributed by AI. 
Any subsequent student edits are overlaid as incremental changes. This differentiation helps instructors not only see the presence of AI involvement but also track how much of the final solution is student-authored versus AI-produced. By linking edits to their source, \sys{} makes authorship transparent without requiring instructors to manually compare logs or chat histories.
\deleted[id=az]{\sys{} also shows messages exchanged between students and AI as a bar chart on top of the timeline, where the blue ones are from students, and gray ones are from students (Figure~\ref{fig:ui-details-ai}.a). Users can hover over to check the exact message content as complementary to the visual patterns of code that originates from AI. }

\subsection{\added[id=az]{Design Probe: }Targeted Question Generation}
\figureQA{}

\replaced[id=az]{We explored whether instructors might benefit from being able to create targeted, learning-oriented questions directly on top of the timeline visualization through this speculative feature.}{\sys{} enables instructors to create targeted, learning-oriented questions directly on top of the timeline visualization.}
By clicking a region of code edits (e.g., where AI suggestions were introduced and not validated, Figure~\ref{fig:ui-qa}.a), \sys{} shows a pop-up window where it could create a context-aware question to assess students' understanding. 
Instructors can \replaced[id=az]{choose between}{select to either create} a multiple-choice question or an open-ended question (Figure~\ref{fig:ui-qa}.b).
The code area shows students' code at the clicked timestamp, highlighting code originated from AI in yellow (Figure~\ref{fig:ui-qa}.c). 
\replaced[id=az]{Instructors could optionally adjust constraints for the LLM prompt (Figure~\ref{fig:ui-qa}.d) to explore how tailored questions might work.}{Instructors could also tailor questions by specifying constraints (Figure~\ref{fig:ui-qa}.d) as part of the prompt sent to the LLM that generates the questions. }

After clicking 'create question' (Figure~\ref{fig:ui-qa}.e), \sys{} generates an example question (Figure~\ref{fig:ui-qa}.f) and an expected answer (Figure~\ref{fig:ui-qa}.g). 
Instructors could further modify \replaced[id=az]{these artifacts}{the question and answers}, and send them to students ((Figure~\ref{fig:ui-qa}.h)).
\replaced[id=az]{This allows us to observe what kinds of questions they valued and how they imagined aligning them with debugging or conceptual-learning goals. This feature was not fully evaluated as part of the system; rather, it served to spark discussion about how question-generation tools might be integrated into timeline-based visualizations.}{This allows them to ask context-specific questions such as “Can you explain what this function is doing?”, grounding the assessment in students’ actual work rather than generic quiz items.
In addition, the flexibility of tailoring questions makes it possible to align questions with both immediate interventions (e.g., debugging misconceptions) and long-term learning goals (e.g., promoting conceptual understanding).}

\subsection{Implementation Details}
\sys{}'s implementation contains two key components:
\begin{enumerate}
    \item A VSCode extension for students, that streams coding history and chat interaction with AI tools, and detects whether code originates from AI or is written by students.
    \item A web application for instructors, including a frontend that visualizes students' AI usage and code edits, allowing instructors to create questions and send them back to students, and a backend server that manages real-time data processing with socket.io and calls LLM APIs to create questions for instructors and forward messages between instructors and students. 
\end{enumerate}

\added[id=az]{Our implementation of \sys{} is open-source}\footnote{Anonymized repository: \url{https://anonymous.4open.science/r/ProTea-A10F}}.

\subsubsection{VSCode Extension for Students.}\label{sec:vscode}
The VSCode extension is built on top of an existing tool that collects keystroke-level edits in the editor from \textit{[redacted for anonymity]}.
We extended it to also capture students' messages exchanged with GitHub Copilot Chat, including the prompt, AI's responses, and the context included when prompting LLMs. 
Every keystroke, edit, and chat will be streamed to the instructor's side website through WebSocket messages. 
It also includes a recording function, where students' edit and chat logs will be recorded and saved in a JSON file. 
VSCode also communicates with the instructor's side website to forward questions from instructors to students, and forward answers from students to instructors. 

\subsubsection{Instructor's Web application}
The frontend visualization is built as a React/TypeScript application that renders (\textit{i}) a timeline view of edit streams with AI attribution highlighting, and (\textit{ii}) a synchronized code view. It supports interactions such as clicking, playback, and question creation. 
The backend is a Node.js server using Socket.IO that manages session state and real-time streaming. The GPT-3.5-turbo model is used for generating questions.

%% file: sections/08-dataCollection.tex
\section{Students' Coding Dataset}

To get a realistic dataset of students' coding histories, we conducted a study to collect students' coding history data on two Python programming problems. This data set is used for evaluating \sys{}. We also published the dataset as a contribution to this paper.

\subsection{Recruitment}

\tableDataCollectionDemographics{}

We reached out to \added[id=az]{both undergraduate and graduate} students over 18 years old from the computer science and information science programs on campus with Python coding experience. 
\replaced[id=az]{Because \sys{} is indented to support visualization of a wide range of AI usage patterns in programming tasks---not only short and simple introductory exercises---we did not restrict participants to novice programmers. Instead, we sought participants with diverse experience so the study could capture a broad spectrum of coding and AI-usage behaviors, including those that arise in more complex codebases.}
{To collect a variety of coding patterns, we had no constraints on their levels of Python coding. }

The study included 20 participants (11 men, 8 women, 1 non-binary) with diverse roles and experience levels. Their programming experience ranged from less than 3 months to 9 years, with most reporting 1--7 years. Participants' roles were varied, including students, software developers/engineers, data professionals (analyst, engineer), a UX researcher, and a technology consultant. This diversity in both experience and professional background provided \replaced[id=az]{a wide range of real-world coding strategies across various levels, containing authentic evidence for evaluating how \sys{} supports identifying AI usage patterns in coding process.}{a broad perspective on coding and AI usage practices.}

\subsection{Method}

Each participant wrote code for two programming tasks in Python. We collected their keystroke coding logs, chat messages with AI (GitHub Copilot in this study), and test results of each run. In total, we collected 20 coding histories for each coding task. We counter-balanced the order of programming tasks they worked on. To ensure we collected a variety of coding patterns and AI usage patterns, we set two different instructions for participants' AI usage when coding. For 10 participants, they used AI without restrictions, and for the other 10 participants, they were asked to only use AI when they were stuck, and not fully rely on it.
\added[id=az]{These conditions were independent of participants’ experience level, as our goal was to elicit varied patterns of AI-usage rather than compare groups by backgrounds.}
The study was conducted virtually on Zoom. Each participant received a \$15 Amazon gift card as compensation.

\subsection{Data Collection Tool}
We built a VSCode\footnote{https://code.visualstudio.com} extension to collect the coding dataset (Section~\ref{sec:vscode}). The extension captures three types of data: (1) keystroke-level coding histories, (2) chat messages with AI (e.g., GitHub Copilot Chat\footnote{https://code.visualstudio.com/docs/copilot/chat/getting-started-chat}), and (3) execution results from test files. After each study session, the collected data is stored in JSON format.
The VSCode extension was deployed on GitHub Codespaces\footnote{https://github.com/features/codespaces}\added{, a cloud-based IDE.}\deleted{ for user study}
\added[id=az]{Participants used the web-based code editor without installing it on their machines}.  
In addition, for each session, we recorded participants' screens as a redundant source of their coding activities and AI interactions. These recordings served as a supplementary resource to validate and proof-check the coding history logs.

\subsection{Tasks}
To ensure our system is capable of visualizing complex code and can generalize to large code samples, we aimed to collect code histories on long and difficult programming tasks. 
We compile tasks from common Python coding problems online that were designed to contain multiple function modules. Both tasks can be solved within 150 to 300 lines of code. 
The tasks were as follows: 


\dgbadge[SENSAICHA]{Task 1 (T1)} A student gradebook system. 
Build a command-line gradebook that contains data in a JSON file, manages students and their course grades, and provides stats on scores. Implement operations to add/list students, add/list grades (with optional filters), and generate reports (per-student grade lists and per-course statistics).

\dgbadge[SENSAICHA]{Task 2 (T2)} A hospital appointment system. 
Build a command-line tool to manage doctors, patients, and appointments with JSON storage. Implement seven actions---add/list doctors and patients, schedule/list appointments, and save \& exit---with strict input validation (unique names, positive rates/durations) and no overlapping appointments per doctor.

We designed two programming tasks of comparable difficulty. Both tasks required participants to work with common data structures such as lists and dictionaries, perform operations on JSON files (load and save), calculate statistics, and apply string formatting.
For each task, participants were provided with starter code containing a task description, function TODOs, and an accompanying test file. The test file allowed participants to run \verb|./test.sh| to check functionality and correctness, providing intermediate feedback during development.
Each task was allotted 20 minutes, and participants were not required to complete the entire task\added[id=az]{, as the goal of the study was not final correctness}. \replaced[id=az]{In order to capture a rich range of coding patterns---including iteration, debugging, prompting, strategy shifts, and AI-assisted edits---participants}{Instead, they} were expected to iteratively develop their solutions starting from the provided starter code and aim to pass the automated checks. \added[id=az]{20 minutes ensured that both novice and experienced programmers could engage meaningfully with the task within the study duration while still producing sufficient interaction data for evaluating \sys{}’s process visualizations.}

\subsection{Results}
We collected 20 coding histories for each Python programming problem. Each coding history lasted 20 minutes. For \dgbadge[SENSAICHA]{T1}, students' solutions ranged from 188 to 319 lines of code\added[id=az]{ ($M=229$, $SD=32.89$)}, with 51 to 1679 keystroke edits\added[id=az]{ ($M=788.40$, $SD=449.78$)}, \deleted[id=az]{from 3 to 39 chats with AI, }0 to 15 runs of the test file. 
\added[id=az]{The number of AI chats per participant ranged from 3 to 39 chats.}
For \dgbadge[SENSAICHA]{T2}, students' solutions ranged from 198 to 306 lines of code\added[id=az]{ ($M=239.25$, $SD=31.11$)}, with 37 to 2610 keystroke edits\added[id=az]{ ($M=769.42$, $SD=611.98$)}, \deleted[id=az]{from 3 to 42 chats with AI, }0 to 15 runs of the test file. 
\added[id=az]{The number of AI chats per participant ranged from 3 to 42 chats.}

\added[id=az]{Across all 40 collected coding histories, 62.67\% of edits were human-authored, 14.91\% were generated via GitHub Copilot autocomplete, 7.25\% were human edits of AI-generated code, and 1.31\% were direct copy-pastes; the remaining edits involved file actions (save/open/close). At the participant level, reliance on AI varied widely: the proportion of AI-contributed edits ranged from 4.47\% to 98.70\% ($M = 23.22\%$, $SD = 21.56\%$), indicating substantial diversity in AI usage patterns across participants, which is expected given the two different AI-usage instructions assigned to participants.}

%% file: sections/05-userStudy.tex
\section{\sys{} User Study}

\sys{} is designed to give instructors better visibility into students’ AI usage during programming and to support targeted guidance when misuse occurs. We conducted a within-subject study to evaluate \sys{}’s effectiveness in monitoring AI usage and guiding interventions. To ensure generalizability beyond short examples, the study used the long-code dataset we collected.
Since no existing tools support monitoring students' AI usage, we built a baseline system for comparison. The baseline was implemented as a code editor paired with students' chat histories with AI, allowing participants to review both code and messages side by side. Participants used both \sys{} and the baseline to answer quiz-style questions about students’ AI usage. This design helped us assess how much the visualization improved instructors' ability to detect and reason about AI misuse, as well as its limitations. The study was reviewed and exempted by the Institutional Review Board (IRB).

\subsection{Method}

\subsubsection{Recruitment}
\tableUserStudyDemographics{}

Because \sys{}'s end users would be programming instructors, we reached out to students and instructors from the authors' universities, who had experience teaching Python programming courses or was at least proficiency in Python. 
During a screening session, participant indicated their prior experience teaching and using Python. 
In total, 12 participants were recruited, including nine men and three women, with Python experience ranging from 4 to 10 years. Most held teaching roles (Graduate Student Instructors, Teaching Assistants, or Tutors), while others included an Engineer, and a graduate student researcher (Table~\ref{tab:user-study-demographics}).

\subsubsection{Baseline System}
\figureBaseline{}

Since no widely used tools exist for monitoring students’ AI usage in programming, we designed our own baseline system (Figure~\ref{fig:baseline}). The baseline presents students’ final code submissions alongside their chat histories with GitHub Copilot Chat for each student.

We designed the baseline this way for several reasons. Chat histories provide direct evidence of how students prompt AI and revise their requests, capturing aspects of interaction that are not visible in code alone. Compared to the visualization in \sys{}, which emphasizes code edits and AI contributions within the coding process, chat histories highlight AI usage from a complementary perspective, revealing the dialogue and reasoning behind code adoption.

This baseline allowed us to investigate how participants perceive different types of information: visualizations versus raw chat logs. In particular, we were interested in the advantages and limitations of each, how users interacted with them, and how future systems might best balance these two sources to support instructors.

\subsubsection{Study Setup}
The study was conducted remotely via Zoom using a within-subjects format where participants used both \sys{} and the baseline system.
We counterbalanced the order of the systems and the tasks (\dgbadge[SENSAICHA]{T1} and \dgbadge[SENSAICHA]{T2}). 
For each condition, we provided 10-20 minutes of training on how to use the system, the programming problems, and how to read the quiz questions. 
After training, participants had 20 minutes to answer quiz questions about students' AI usage using the assigned system. 
\added[id=az]{The quiz questions required participants to analyze students' coding histories and identify concrete aspects of AI usage. For example, which functions were AI-generated or edited, which students relied most on AI, and how students’ strategies changed over time (e.g., starting manually then switching to AI). These questions assess the task of inferring AI involvement and authorship from coding histories.}
After finishing the quiz questions for each system, participants were asked to complete a survey about their experience using the system. At the end of each study, we conducted a reflective survey and interview to compare the two systems. 
We encouraged participants to ask any questions about the usage of both systems. 
Each study lasted about 70 minutes. 

\subsubsection{Data Collection and Metrics}
During the study, we recorded participants’ screens as they completed the tasks, capturing their quiz responses, think-aloud audio, and interactions with both systems. We also collected their answers to the post-study survey and follow-up interviews. Two members of the research team were present during each session.
To evaluate participants' quiz performance, we calculated the F-score as accuracy for each multiple-choice question as:
$$\frac{2*True\ Positive}{2*True\ Positive + False\ Positive + False\ Negative}$$
For single-choice questions, accuracy was binary (1 if correct, 0 otherwise). Open-ended quiz responses were coded independently by three authors, who collaboratively developed a coding scheme to ensure reliability.
We also coded participants’ screen recordings to analyze how they interacted with each system, including time spent on quiz questions and strategies used to extract relevant information. A thematic analysis was conducted to identify recurring patterns and insights into participants’ survey and interview answers.
Finally, a paired t-test was used to compare participants’ performance \added[id=az]{(quiz accuracy)} and workload \added[id=az]{(time spent answering quiz questions)} across the two systems. 
\added[id=az]{Normality was verified for both accuracy and time (Shapiro–Wilk tests, $p>0.05$), supporting the use of the paired t-test.} 
Figure~\ref{fig:comparison-survey} presents the results of the comparison survey\added[id=az]{, in which participants explicitly compared the two conditions.}.

\figureComparisonSurvey{}

\tableSystemSurvey{}

\subsection{Results}

\subsubsection{Participants understand students' AI usage more accurately using \sys{} than the baseline.}

We first compared participants’ accuracy in answering quiz questions, which required them to identify students’ AI usage, how AI-generated code was adapted, and how much of the final code was written by students. 
\added[id=az]{Accuracy was calculated as the proportion of correct answers for each participant, averaged across multiple-choice (F-score) and single-choice (binary) items to produce an overall accuracy score.}
Accuracy was significantly higher with \sys{} ($M=0.79$, $SD=0.13$) than with the baseline system ($M=0.31$, $SD=0.14$, $p<0.00001$).
This quantitative result aligns with findings from participants’ self-report survey (Table~\ref{tab:outcomes}) and the comparison survey (Figure~\ref{fig:comparison-survey}). Overall, participants reported that \sys{} was more helpful in (1) understanding AI usage patterns, (2) identifying students who needed intervention, and (3) detecting unproductive usage that conflicted with pedagogical goals.

Interestingly, however, the perceived difference between \sys{} and the baseline was smaller in self-reports than in the quiz results. \replaced[id=az]{Although self-reported understanding did not differ significantly across conditions (Table~\ref{tab:outcomes}), objective quiz accuracy showed a significant improvement when using \sys{}.}{When using the baseline, participants believed they had a better understanding of students’ AI usage than they actually demonstrated in the quiz.} This discrepancy suggests that raw code and chat histories may create a false sense of confidence, whereas the visualization in \sys{} provides more accurate and actionable insights. We further discuss these findings with qualitative evidence in Section~\ref{sec:perception-gap}.

\subsubsection{\sys{} takes less effort to identify unproductive AI usage patterns.}
To assess whether \sys{} takes less effort to understand students' AI usage and identify unproductive AI usage patterns, we measured the time spent answering the quiz questions. 
Participants took \added[id=az]{slightly} longer in the baseline condition ($M=841.75$ seconds, $SD=223.18$) compared to the \sys{} condition ($M=815.75$ seconds, $SD=221.75$). 
However, we did not find a statistically significant difference ($p>0.05$).

We concluded several reasons for this result between the two conditions. 
First, participants mentioned \sys{} requires some learning curve to effectively use the system, while in the baseline, the chat messages are very easy to understand and clearly show whether students ask to explain concepts or simply ask for a solution. 
Second, in the baseline condition, participants had to manually read through the chat history, and some of them found it very tedious and gave up midway due to the large amount of text, and quickly moved to the next question.

\subsection{System Usability and Study Insights}
\subsubsection{\sys{} helps participants more easily differentiate productive and unproductive AI usage patterns. }
Participants find the visualization in \sys{} could help them quickly understand how students use AI in their code, such as when they used it, which part of the code they used it, how they adapt AI-generated code, and their editing patterns on AI-generated content. 
\mylongquote{The timeline and getting to see all the indicators of where the students' activity was flagged and what exact sections were flagged as copy-paste helped see whether students needed intervention and also whether students were actually copy-pasting or whether it was just a small section of code that could only be written in one way. (P12)}

In addition to knowing whether the code edit is from AI, participants could also see how students were engaged with the AI-generated code. 

\mylongquote{I would see if there are any edits to the code after an AI copy-paste or AI-generated flagging. This indicates that although they are using AI as a foundation, they are still making their own edits on top of the content, showing that they are still using critical thinking and adding their own work to the code.
I would also look at the quantity of the lines and how they are spread out. If they are all over the code, it seems like the student has used it to do all the work, while if I see only a few lines in use and they are all localized, it seems like the student has gotten stuck on one spot and used AI to help them. (P12)}

\subsubsection{When using the baseline, participants believed they had a better understanding of students’ AI usage than they actually demonstrated.}\label{sec:perception-gap}
While participants showed significantly lower quiz accuracy with the baseline, several participants reported that they had a good understanding of AI usage patterns using baseline \added[id=az]{and no statistical significance found in their self-report results (Table~\ref{tab:outcomes})}. 
We identified the following reasons for this gap between their actual and perceived understanding of AI usage.
First, participants thought they understood well, as they took the chat histories as direct evidence of how students prompt AI. For example, they can know that if students are asking AI to explain a concept or simply send the entire problem description to get a solution. 
\mylongquote{Seeing how the student is prompting the AI was really helpful. If the student was just copy-pasting the code instructions into the AI and copying the code back into the editor, then it's clear that their use of AI is unproductive. However, other students wrote some code and were asking the AI to help with syntax, or were asking the AI to help explain the prompt, or were asking AI how to express a small thing, which was more productive for their learning. (P12)}

While chat history is also provided in \sys{}, it is not prominent on the timeline view, and participants have very little use of it during the study. 
Participants commented that the timeline visualization and the chat histories provide two different aspects of AI usage, and could be more helpful if combined to provide insights into students' behavior (P10). 


Participants had two main different interaction patterns when using the baseline. They either go through each message to see how students draft the prompts and how AI responded, or they simply skim through the messages and put a random answer in the quiz. 
We found two reasons for the latter pattern. First, participants do not have the bandwidth to read through the text-heavy conversations.
\mylongquote{Very hard to tell [students' AI usage]. The system is only listing all the chat history. Hard to distill insightful points. (P5)}

Second, participants cannot easily see the connection between the text messages and the way students adapt them to their code. 

\mylongquote{The color area and the code allocation is more easier to identify exactly, the code segment that students are trying to use AI or trying to do more modifications. It's easier to see this kind of comparison. But the first system [baseline]. I need to compare them, like, manually by my own understanding, it's very difficult to use that, actually. (P6)}

\subsubsection{\sys{} helps to identify the timing and students who need intervention. }
Most participants agree that \sys{} is more helpful in identifying the timing and students who need intervention. 
With \sys{}, participants can easily see at what time students copy-paste, autocomplete, or even type in code that is very similar to AI-generated code. 
This could help participants decide when to intervene once they see unproductive AI usage patterns. 
\mylongquote{[...] the colored lines which indicate whether AI was used, or the student's code looks like AI, or there was copy-paste [...] gives a good visual to see what is going on. Where the student is using AI the most, and at which part of the code. (P9)}

Participants could also quickly intervene through the visualization. By quickly seeing parts that are contributed by AI in students' code, they could directly interact with the visualization to intervene, such as giving hints, asking students about their understanding, and sending quizzes. 
\mylongquote{I would use this tool to find which students are using AI in an unproductive way and make sure to talk to them about their usage and make a plan to avoid using it as much. I would also use the question function to test the student to check if they actually know the content. Checking how much knowledge the student has can help determine the path to assist the student. (P8)}

\subsubsection{Even with the same visualization, participants may disagree on what constitutes ``productive'' v.s.  ``unproductive'' usage.}

Participants had diverse opinions on what counts as productive versus unproductive AI use, even when looking at the same visualization patterns. Some considered heavy reliance on AI (entire blocks of AI-generated code) productive, as they interpret the student being able to assemble a working solution. Others saw that same pattern as unproductive, since they think it suggested minimal learning and thinking. Similarly, some viewed back-and-forth edits on AI-generated code as signs of good learning, where students validate and explore the AI-generated code, while others saw such edits as evidence of struggle and inefficiency. Overall, judgments of productivity varied based on how instructors interpreted the role of AI in the learning process.

This finding aligns with what we found in the formative study (Section~\ref{sec:formative-variation}). There is a wide variation in how students use AI. Such wide variation leads to different understandings of whether the usage is productive or not. Such disagreement on the usage patterns also points to the need to quickly assess students' understanding of specific concepts. 

\subsection{Preliminary Evaluation on Targeted Question Creation}
\replaced[id=az]{As part of our early exploration}{We design the visualization of \sys{} to support further interactions for instructors to guide students' AI usage and intervene in a timely manner. In the current design}, we implemented the QA function as a proof-of-concept, \replaced[id=az]{to probe how instructors might imagine using timeline interactions to assess students’ understanding}{one of the ways instructors could achieve this goal}. While we did not evaluate this function fully in a deployed study with students, we \deleted[id=az]{asked participants to play with the QA feature and }gather preliminary feedback from participants on the usefulness of this feature, also providing insights into future design considerations.

Overall, participants had mixed opinions on creating questions from the visualizations. 
Some participants find the feature to be a helpful way to test if the student has fully understood the AI-generated content (P2, P5, P8, P11, P12). 
\added[id=az]{P2 mentioned that \myquote{when we see that a student is not able to understand the concept, we are able to send out the relevant resources to them regarding implementing that particular function.}}

\mylongquote{\deleted[id=az]{Yes, it's very helpful. When we see that a student is not able to understand the concept, we are able to send out the relevant resources to them regarding implementing that particular function. If they are able to learn the concept, then next time they will be able to implement it on their own. It would be great. I find it useful. (P2)}}

On the other hand, participants also brought up concerns about the QA interaction. 
\replaced[id=az]{First, P12 raised concerns about relying on students’ answers to accurately infer understanding, noting confounding factors such as misreading, guessing, or differing interpretations.}{First, participants were not sure that even with a QA activity, what signals they could use from students' answers to accurately assess their understanding.}
\mylongquote{\deleted[id=az]{[...] there are too many confounding variables that impact my ability to make a decision based on the student's answer to the question. It could be that the student didn't know the answer because they were using AI, but it could also be that they read the question wrong, misclicked, misinterpreted what I wanted to ask, or just think about things a different way than I do. I think the feature is cool, but it's hard to derive a conclusion from it about students' AI usage. (P12)}}

Second, participants were suspicious about the quality of the generated questions. Specifically, they were concerned about (1) if the questions truly targeted the intended code regions (P1, P12), (2) if the question's difficulty was appropriate (P6, P9), and (3) how much effort would be required to refind the questions (P4). 

\mylongquote{\deleted[id=az]{I think multiple choice questions are easy to answer, and open-ended questions might be sometimes difficult to answer by students, as they are more system design rather than code-specific, probably. Like the potential risk of implementing / not implementing a function, rather than just talking about syntax-wise or other types of help. (P9)}}

\replaced[id=az]{P10 mentioned that in real classrooms they would rather address misunderstandings directly through instruction than send additional questions, and doubted that such questions would meaningfully distinguish between genuine understanding and AI over-reliance.}{P10 thinks the QA feature would not be needed in a real classroom because they prefer to directly incorporate it into their instruction, rather than sending additional questions. They also think it is rare that a student overly relies on AI but can answer quiz questions successfully. }
\mylongquote{\deleted[id=az]{[...] I just felt that it was unnecessary because I can totally see that for Participant 20, they literally copied and pasted the entire assignment themselves.	
So even if they were, like, going to understand what the AI generated for them, like, each function, and go, like, understand what the AI is saying.	
I mean, that would be, like, a 1\% case, honestly.	
Because 99\% of the time, it would mostly be that the person has copied and pasted because they themselves don't know it. (P10)}}

%% file: sections/06-discussion.tex
\section{Discussion}

\subsection{Implications for Learning Programming with AI}

\subsubsection{Bridging gaps in visibility. }
A key contribution of \sys{} is that it closes the visibility gap between what students do with AI and what instructors are able to observe. Traditional approaches, such as reviewing only the final code or relying on students’ self-reports, often obscure the underlying process of prompting AI, accepting or rejecting suggestions, and iteratively editing. Furthermore, as our studies showed, even providing direct evidence such as chat history, instructors still have little understanding of how AI-generated code is adapted. 
By surfacing when AI was invoked, where in the code it was used, and how students integrated AI-generated content, \sys{} provides instructors with a clearer picture of the learning process. This level of transparency transforms what was previously a “black box” into a structured narrative, enabling instructors to distinguish between productive uses of AI (e.g., scaffolding or debugging) and problematic ones (e.g., blind copy–paste or over-reliance).

\subsubsection{Pedagogical value. }
Our preliminary findings showed that, with greater visibility, instructors can better align their interventions with their pedagogical philosophies. For those who adopt a more cautious strategy towards AI, \sys{} provides the evidence needed to intervene when students bypass essential learning steps. For instructors who encourage exploration with AI, the system offers a way to ensure that students still engage meaningfully with concepts rather than deferring all reasoning to the model. In both cases, \sys{} demonstrated the potential capability of enabling instructors to personalize feedback. For example, by encouraging a struggling student to slow down and validate AI output, or by supporting advanced students in using AI to explore alternative approaches. Instead of one-size-fits-all advice, instructors can tailor their guidance to individual students’ behaviors and goals.

\subsubsection{Policy-making support.}
Another implication is that \sys{} can inform policy development at the course or institutional level. As we found in formative studies, many current AI usage policies are binary because instructors lack the means to monitor nuanced interactions (Section~\ref{sec:formative-policy}). By providing a fine-grained record of how AI is used in context, \sys{} creates the foundation for more flexible and context-aware policies. For instance, policies could distinguish between acceptable uses of AI for a small part of help versus unacceptable uses for generating entire solutions. Over time, aggregated data from systems like \sys{} could reveal broader patterns across cohorts, helping educators refine policies that balance academic integrity with the pedagogical benefits of AI.

\replaced[id=az]{Overall, by making AI use more visible, aligning feedback with teaching goals, and informing more nuanced policy considerations, \sys{} can contribute to how programming instructors respond to and integrate generative AI in CSEd.}{Overall, by making AI use transparent, aligning feedback with teaching philosophies, and supporting more nuanced policy-making, \sys{} has the potential to fundamentally reshape how programming instructors adapt to generative AI in CSEd.}


\subsection{\added[id=az]{Ethical Implications and Privacy in \sys{}}}\label{sec:disc-privacy}

\added[id=az]{
As \sys{} tracks students’ coding progress and interactions with AI tools---such as GitHub Copilot and ChatGPT---it raises potential privacy concerns, including monitoring activities outside class assignments. Students may also be reluctant to share their coding process, and sensitive information could reinforce stereotypes or bias in teaching~\cite{medel2017eliminating, robinson2018using}.
To address these concerns, \sys{} allows students to disable edit history and AI usage tracking at any time in VS Code. To prevent private issues, it can be deployed locally using private cloud services or LLM APIs, and configured to access only in-editor AI interactions.
In this paper, we focus on instructors’ perspectives. Future work could examine students’ experiences through a classroom deployment and explore features such as in-editor notifications that inform students when monitoring is active.
}

\subsection{Limitation and Future Work}

Our work has the following limitations on the system itself and the study design. 
First, we aim to design this tool as a way for instructors to spot unproductive AI usage and provide personalized guidance to students, but only evaluate with instructors, without students. It would be valuable for future work to explore how the tool applies in real classrooms. 
\added[id=az]{our study examined only two programming tasks, both implemented in Python and of a similar level of complexity. As a result, the generalizability of our findings to other task types, difficulty levels, and programming languages is limited. Future work should evaluate \sys{} across a broader range of programming contexts.}
Our preliminary results on the usefulness of the QA feature and potential improvements provide insights into future design considerations. Future work could explore designs that could not only easily assess students' understanding, but also encode useful metrics in the questions to make students' thought processes more visible to instructors. The current study only involved 12 participants from the instructor's perspective with controlled tasks. Larger and in situ classroom deployment studies could reveal new and realistic challenges. In addition, broader AI ecosystems (e.g., ChatGPT, Cursor AI, Codex) may also differ in students' strategies. 

Second, \sys{} could be extended to deliver ``better information'' to instructors. Now \sys{} only shows how AI contributed to the code, not embedding students' understanding in the visualization. With additional functionalities to assess students' understanding, the visualization could be extended to provide insights into how students' mental models change. For example, instructors might see a student rely heavily on AI, not able to correctly answer questions, gradually getting more understanding of the concepts, and gradually taking over the AI-contributed code by modifying and verifying the AI code. 

Third, we found that even with the same visualizations, instructors may disagree on what constitutes ``productive'' vs. ``unproductive'' usage. Future work could also explore if there are any standard ways to classify good and bad usage of AI, or supporting instructors to customize their definition of good and bad AI usage, and customize their own metrics to spot it.

\added[id=az]{While \sys{} currently focuses on visualizing each individual student’s coding history, an important next step is supporting aggregated views of AI usage across students and assignments. Such analyses would help instructors identify broader patterns. For example, whether many students relied on AI for loop syntax, certain sub-tasks, or specific code transformations. Future work could extend \sys{} to surface recurring AI-generated snippets, common intervention points, or shared misconceptions across the class. These aggregated visualizations would complement the individual-level views by enabling instructors to understand classroom-wide trends and design targeted teaching interventions.
}

%% file: sections/07-conclusion.tex
\section{Conclusion}
In this work, we conducted a survey study to examine instructors' current practices, pain points, and needs around understanding students’ AI usage in programming and providing guidance. We introduced \sys{} to address a major challenge identified in the survey: the visibility gap between how students actually interact with AI and what instructors can observe from final submissions. \sys{} bridges this gap by visualizing AI usage as trails in students’ code edits, enabling instructors to distinguish productive from unproductive patterns, create in-situ assessments, and align AI use with pedagogical goals. Our evaluation showed that \sys{} allowed instructors to identify AI usage with twice the accuracy of a baseline system and to more effectively determine when and with whom to intervene. By highlighting how students adapt AI-generated code, the visualization also provides rich meta-information about students’ problem-solving approaches, reducing the effort required to understand text-heavy AI dialogues. Finally, this work offers design insights for facilitating feedback loops among instructors, students, and AI.
